\documentclass[prl,reprint,aps,superscriptaddress,longbibliography,floatfix]{revtex4-2}
\usepackage{graphicx}
\usepackage{bm}
\usepackage{amsmath}
\usepackage{amssymb}
\usepackage{booktabs}
\usepackage{multirow}
\usepackage{enumerate}
\usepackage{mhchem}
\usepackage{upgreek}
\usepackage[dvipsnames]{xcolor}
\usepackage{comment}

\usepackage[normalem]{ulem} 
\usepackage[hidelinks]{hyperref}
\hypersetup{colorlinks=true, linkcolor=blue, citecolor=red, urlcolor=blue}

\newcommand{\lp}{\left}
\newcommand{\rp}{\right}

\newcommand{\ket}[1]{|{#1}\rangle}

\newcommand{\bra}[1]{\langle{#1}|}

\newcommand{\0}{{\bm 0}}
\newcommand{\AAA}{{\bm A}}
\newcommand{\aaa}{{\bm a}}

\newcommand{\dd}{{\bm d}}

\newcommand{\GG}{{\bm G}}
\newcommand{\gggg}{{\bm g}}
\newcommand{\kk}{{\bm k}}
\newcommand{\nn}{{\bm n}}
\newcommand{\pp}{{\bm p}}
\newcommand{\qq}{{\bm q}}

\newcommand{\rr}{{\bm r}}

\newcommand{\uu}{{\bm u}}

\newcommand{\nnabla}{{\bm\nabla}}

\newcommand{\up}{\uparrow}
\newcommand{\down}{\downarrow}

\newcommand{\mcM}{{\mathcal{M}}}

\newcommand{\rmH}{{\rm H}}
\newcommand{\rmF}{{\rm F}}
\newcommand{\rmM}{{\rm M}}
\newcommand{\rmm}{{\rm m}}
\newcommand{\rmX}{{\rm X}}

\begin{document}

\title{Moir\'e-induced altermagnetism from nonmagnetic constituents}

\author{Jingtian Shi}
\affiliation{Materials Science Division, Argonne National Laboratory, Lemont, Illinois 60439, USA}

\author{Maxim Khodas}
\affiliation{Racah Institute of Physics, Hebrew University of Jerusalem, Jerusalem 91904, Israel}
\affiliation{Materials Science Division, Argonne National Laboratory, Lemont, Illinois 60439, USA}

\author{Ivar Martin}
\affiliation{Materials Science Division, Argonne National Laboratory, Lemont, Illinois 60439, USA}

\date{\today}

\begin{abstract}

We propose a mechanism for nonmagnetic materials to develop altermagnetic order by moir\'e interference with nonmagnetic substrate, which is driven by structural relaxation and spontaneous twirls in moir\'e domain walls of lattice-mismatched moir\'e square lattices. When doped with one electron per moir\'e domain, a correlated insulating gap is opened by electron interaction. Depending on the location of the moir\'e potential minima, the moir\'e bands can show $d$-wave or $g$-wave altermagnetic splitting. The former can be enhanced by a finite twist angle; the latter is sensitive to strains that drive a transition to $d$-wave.

\end{abstract}

\maketitle

\paragraph{Introduction}

Altermagnetism is a recently identified class of compensated magnetic states in which opposite spin sectors are related by rotation and/or reflection \cite{smejkal2022conventional, smejkal2022emerging, song2025altermagnets}, permitting momentum-dependent spin splitting of electronic bands with symmetry-protected touching points or nodal lines.
Such types of band structure enable a range of exotic features, including spin-polarized currents \cite{jungwirth2026altermagnetic, dou2025anisotropic, chen2025helicitycontrolled} and anomalous Hall effects \cite{smejkal2020crystal, smejkal2022anomalous, attias2024intrinsic, han2026magnetizationfree} under zero net magnetization, which are promising ingredients for integrated electrical control of stray-field-free magnetic memory \cite{jungwirth2026altermagnetic, han2024electrical, chen2025electrical};
besides, both metallic and insulating altermagnets can display piezomagnetism without spin-orbit coupling (SOC) \cite{ma2021multifunctional, naka2025nonrelativistic, karetta2025straincontrolled, khodas2026tuning}.
The richness of altermagnetic phenomena prompted an active search for material realizations \cite{lee2024broken, krempasky2024altermagnetic, reimers2024direct, ding2024large, jiang2025metallic, zhang2025crystalsymmetrypaired, reichlova2024observation, zhang2026arpes}. However, crystalline atomic materials offer limited flexibility and tunability across the broad landscape of correlated physics. This calls for complementary routes to engineer altermagnetism in top-down designed artificial materials.

Moir\'e materials have been emerging as versatile platforms for engineering exotic phases of matter thanks to their exceptional tunability \cite{andrei2021marvels}.
In van der Waals bilayers, a small twist angle or lattice mismatch produces a long-period moir\'e pattern that flattens the electronic bands and enhances electron interaction effects \cite{bistritzer2011moire}. The resulting states include correlated insulators \cite{cao2018correlated}, superconductivity \cite{cao2018unconventional, xia2025superconductivity, guo2025superconductivity}, Wigner crystals \cite{regan2020mott, xu2020correlated, huang2021correlated}, and integer \cite{serlin2020intrinsic, li2021quantum} and fractional quantum anomalous Hall states \cite{cai2023signatures, zeng2023thermodynamic, park2023observation, xu2023observation, lu2024fractional}.
The advancement of moir\'e engineering has motivated several designs of tunable altermagnetism via van der Waals coupling of \textit{magnetic} constituents \cite{mellado2025magnetic, sheoran2024nonrelativistic, liu2024twisted, guo2024valley, sheng2025ubiquitous, zhao2025ferroelectricitydriven, pathak2026strain, ruiz2026twistinduced, ruiz2026emergent, liu2026sublayerresolved, cui2026altermagnetic}.
Given that both ferromagnetism \cite{sharpe2019emergent} and antiferromagnetism \cite{tang2020simulation, xia2026bandwidthtuned} have been realized with moir\'e stacking of \textit{nonmagnetic} materials, a natural question is whether altermagnetism can be engineered in similar ways.

In this work, we report a minimal model study of moir\'e material consisting of \textit{nonmagnetic} layers, which predicts emergent altermagnetism.
Importantly, we engineer the altermagnetic symmetry by utilizing not only the electronic, but also the structural versatility of moir\'e materials --
our work builds on discovery of $\sqrt{2} \times \sqrt{2}$ order in spontaneous twirls \cite{shi2026spontaneous} in lattice-relaxed domain walls of square-lattice moir\'e systems, demonstrating that emergent structural orders in moir\'e materials can give rise to novel electron orders.

\paragraph{Summary of mechanism}

\begin{figure}[t]
	\centering
	\includegraphics[width=\columnwidth]{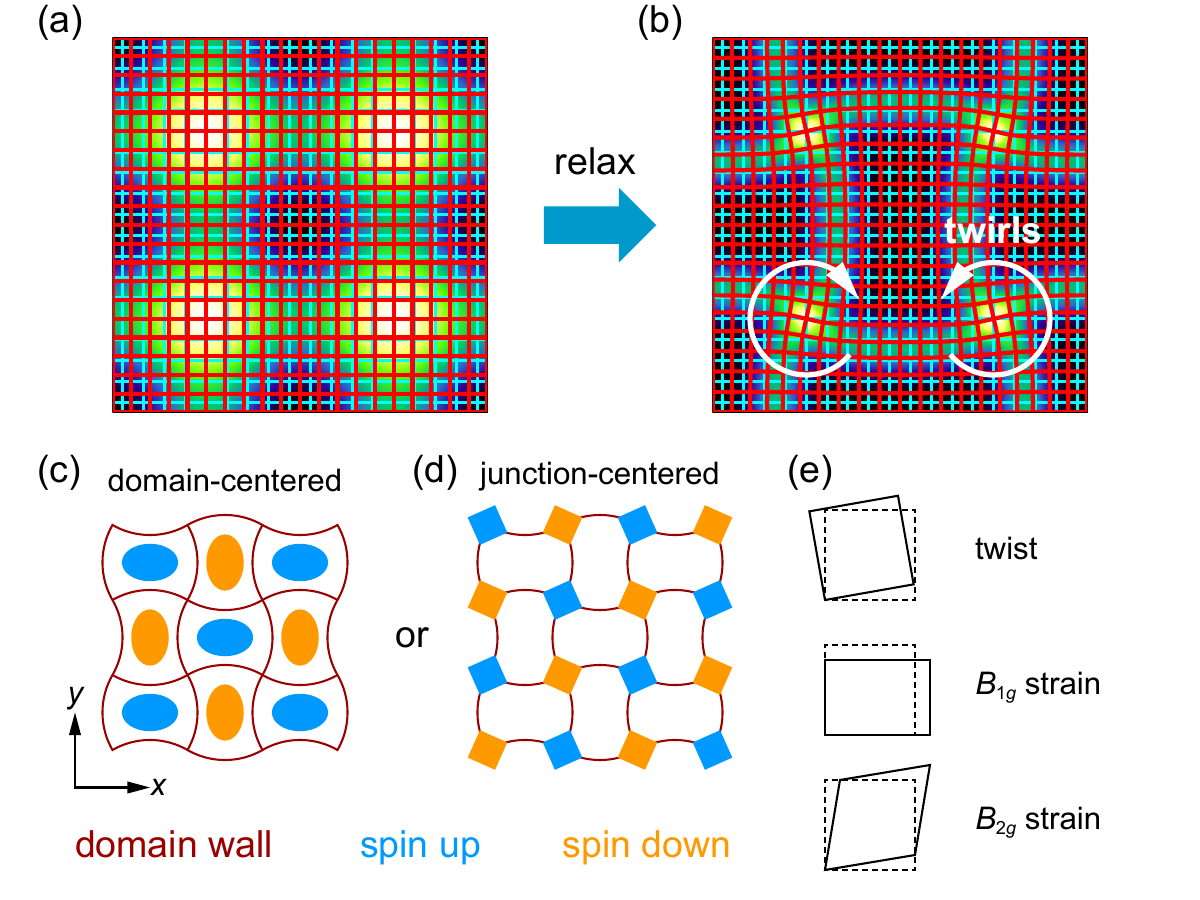}
	\caption{(a), (b) Schematic of lattice relaxation of lattice-mismatched moir\'e square lattice from (a) the rigid configuration to (b) the stable configuration. The red and cyan grids respectively represent an elastic material layer and a rigid substrate, the former having a smaller natural lattice constant than the latter. 
    In the background, darker (brighter) color represents regions with lower (higher) local interlayer coupling energy, or the moir\'e domain (domain wall) regions. The curved arrows mark the twirl chiralities. (c), (d) Schematics of Wannier orbitals of electron moir\'e bands in the two scenarios described on the top of each panel. The coordinate frame is marked at the bottom left of (c). The dark red lines represent the domain walls, and the blue and orange shapes represent occupied electron orbitals with different spin species, as illustrated on the bottom. (e) Schematics of twist and two types of strains applied to the elastic layer we consider in our work.}
	\label{fig:intro}
\end{figure}

As Figs. \ref{fig:intro} (a), (b) show, moir\'e lattice relaxation expands regions with energetically favorable local interlayer stacking registry to domains, and shrinks other regions into domain walls \cite{carr2018relaxation, rosenberger2020twist, weston2020atomic, nam2017lattice}. The domain wall junctions further save energy by twirling, a phenomenon that can also occur in triangular moir\'e domain structures \cite{dai2016twisted, quan2018tunable, zhu2020modeling, maity2021reconstruction, mesple2023giant, kaliteevski2023twirling, shi2026spontaneous, yan2026intertwined}.
In the lattice-mismatched square lattice case relevant to our work, neighboring twirls tend to have opposite chirality, forming a staggered pattern in the square lattice (Fig. \ref{fig:intro}(b)). When the relaxed material is a semiconductor or point-node semimetal, the periodic moir\'e potential splits the electron bands into moir\'e bands \cite{bistritzer2011moire, wu2019topological}, whose Wannier centers depend on the locations of moir\'e potential minima. Two cases are schematically shown in Figs. \ref{fig:intro} (c), (d). 
Electron interaction in the bands gives rise to (extended) Hubbard model physics \cite{wu2018hubbard, shi2026$g$valley} that generates both antiferromagnetic superexchange and ferromagnetic direct exchange couplings between electron spins of neighboring orbitals \cite{anderson1959new, macdonald1988tu, hu2026ferromagnetism}.
When the antiferromagnetic coupling dominates,
because of the symmetry breaking by twirls, the system shows altermagnetic symmetry rather than conventional antiferromagnetic symmetry \cite{das2024realizing, ferrari2024altermagnetism, kaushal2025altermagnetism}. 
Assuming collinear spins and absence of SOC, we identify two cases: i) \textit{domain-centered orbitals} (Fig. \ref{fig:intro}(c)), where the system has symmetries $[C_{2\perp}{\parallel}C_{4z}]$, $[E{\parallel}M_{\hat x}]$ and $[E{\parallel}M_{\hat y}]$,
realizing the spin Laue group ${}^24/{}^{1}m^2m{}^1m$ (notation adopted from Fig. 2 of Ref. \cite{smejkal2022conventional}) with $d_{x^2-y^2}$-wave spin splitting of moir\'e bands;
ii) \textit{junction-centered orbitals} (Fig. \ref{fig:intro}(d)), where the system has symmetries $[E{\parallel}C_{4z}]$, $[C_{2\perp}{\parallel}M_{\hat x}]$ and $[C_{2\perp}{\parallel}M_{\hat y}]$,
realizing the spin Laue group ${}^14/{}^{1}m^2m{}^2m$ with $g_{xy(x^2-y^2)}$-wave spin splitting of moir\'e bands.
Here we have adopted the standard convention for the symmetry operation notation $[A{\parallel}B]$, where the spin (orbital) operation is on the left (right) \cite{smejkal2022conventional}.
$C_{2\perp}$ denotes spin flip, $E$ denotes identity in spin sector, $C_{4z}$ denotes 4-fold in-plane counterclockwise spatial rotation of orbitals about a domain junction point, and $M_\nn$ is spatial reflection of orbitals about a mirror that passes a domain center and is normal to vector $\nn$.
In both cases, the system also has the symmetry of spin-flipping diagonal reflection combined with moir\'e lattice translation: $[C_{2\perp} {\parallel} M_{\hat{x}+\hat{y}} | \tau_m]$ and $[C_{2\perp} {\parallel} M_{\hat{x}-\hat{y}} | \tau_m]$, which are products of the four-fold rotation and normal-mirror reflection symmetries.
$\tau_m$ is the primitive translation of the pre-relaxation moir\'e, \textit{i.e.}, half diagonal translation of the relaxed $\sqrt{2}\times\sqrt{2}$ superlattice.

We will show in the rest of this paper that when the effective electron mass of the parent semiconductor band is relatively small, the $d$-wave spin splitting in the domain-centered case can reach experimentally detectable range, whereas the $g$-wave spin splitting in the junction-centered case is expected to be rather weak. Before elaborating on our model settings, we note  that the altermagnetic order is tunable with twist angle and strain engineering. Schematics of twist angle and strains are shown in Fig. \ref{fig:intro}(e). In the domain-centered case, a finite twist angle between the mismatched layers breaks all the mirror symmetries but preserves $[C_{2\perp}{\parallel}C_{4z}]$,
reducing the ${}^24/{}^{1}m^2m{}^1m$ spin Laue group to ${}^2 4/{}^1 m$ with the $d$-wave spin splitting that is a mixture of  $d_{x^2-y^2}$- and $d_{xy}$-wave components.
We will see that such a twist can enhance altermagnetic band splitting and related properties.
In the junction-centered case, a $B_{1g}$ ($B_{2g}$) strain exerted on one layer breaks $[E{\parallel}C_{4z}]$ and $[C_{2\perp} {\parallel} M_{\hat{x}\pm\hat{y}} | \tau_m]$ ($[E{\parallel}C_{4z}]$, $[C_{2\perp} {\parallel} M_{\hat x}]$ and $[C_{2\perp} {\parallel} M_{\hat y}]$) while preserving the remaining spin-flip reflection symmetries.
The ${}^14/{}^{1}m^2m{}^2m$ spin Laue group hence reduces to $^{2}m^{2}m^{1}m$ with $d_{xy}$- or $d_{x^2-y^2}$- wave band splitting in the $B_{1g}$ or $B_{2g}$ case, respectively. 
In the End Matter, we will see that for both types of strains the $d$-wave order begins to dominate over the original $g$-wave order from very tiny strain strengths.

\smallskip
\paragraph{Model.}

Our model consists of lattice relaxation part and electron part. For lattice, we use the continuum elasticity formulation \cite{san-jose2014spontaneous, nam2017lattice, bennett2022theory, shi2026spontaneous} of an elastic layer on rigid substrate, which minimizes the sum of elastic and substrate coupling energies by varying the in-plane local displacement function $\uu(\rr) = (u_x(\rr), u_y(\rr))$ under moir\'e super-periodic constraints defined by the lattice constant mismatch $\epsilon$, the twist angle $\theta$, and the $B_{1g}$/orthorhombic and $B_{2g}$/shear
strains (see Fig. \ref{fig:intro}(e)) exerted to the elastic layer prior to relaxation. Throughout this work we let the natural lattice constant of the elastic material be $\epsilon = 5\%$ larger than the substrate and allow $\theta$, $B_{1g}$ and $B_{2g}$ to vary. 
Systems with other values of $\epsilon$, including negative values, are related via rescaling of model parameters \cite{SM2}.

We characterize the stiffness of the elastic material with three modulus coefficients $\lambda$, $\mu$ and $\mu_s$, the former two being the Lam\'e coefficients, and $\mu_s$ an independent component of shear modulus allowed by the symmetry of square lattices. See Supplemental Material \cite{SM2} for detailed definitions. Ratios between $\lambda$, $\mu$ and $\mu_s$ vary widely across materials \cite{chen2020twodimensional, chen2021structural, lv2020metallic, liu2023computational}. In major part of this work we take the isotropic limit $\mu_s = \mu$ and the 2D Cauchy relation limit $\lambda = \mu$. The coupling strength to the substrate is characterized by a single real parameter $V_1$, which is the first-star Fourier coefficient of the 
coupling energy density $V(\dd)$ \cite{shi2026spontaneous, SM2} as a function of the local stacking registry $\dd = \dd(\rr) = \mcM\rr + \uu(\rr)$. Here $\mcM\rr$ is the spatially linear pre-relaxation local stacking registry caused by $\epsilon$, $\theta$, $B_{1g}$ and/or $B_{2g}$.
Domain wall twirling only happens when $V_1/\mu$ is above a threshold that is at the order of $\epsilon^2$ \cite{shi2026spontaneous}, and the $\sqrt{2} \times \sqrt{2}$ order in twirls illustrated in Fig. \ref{fig:intro}(b) is only stable when $V_1/\mu$ is below a larger threshold, above which further breaking of periodicity occurs.

\begin{figure}
	\centering
	\includegraphics[width=\columnwidth]{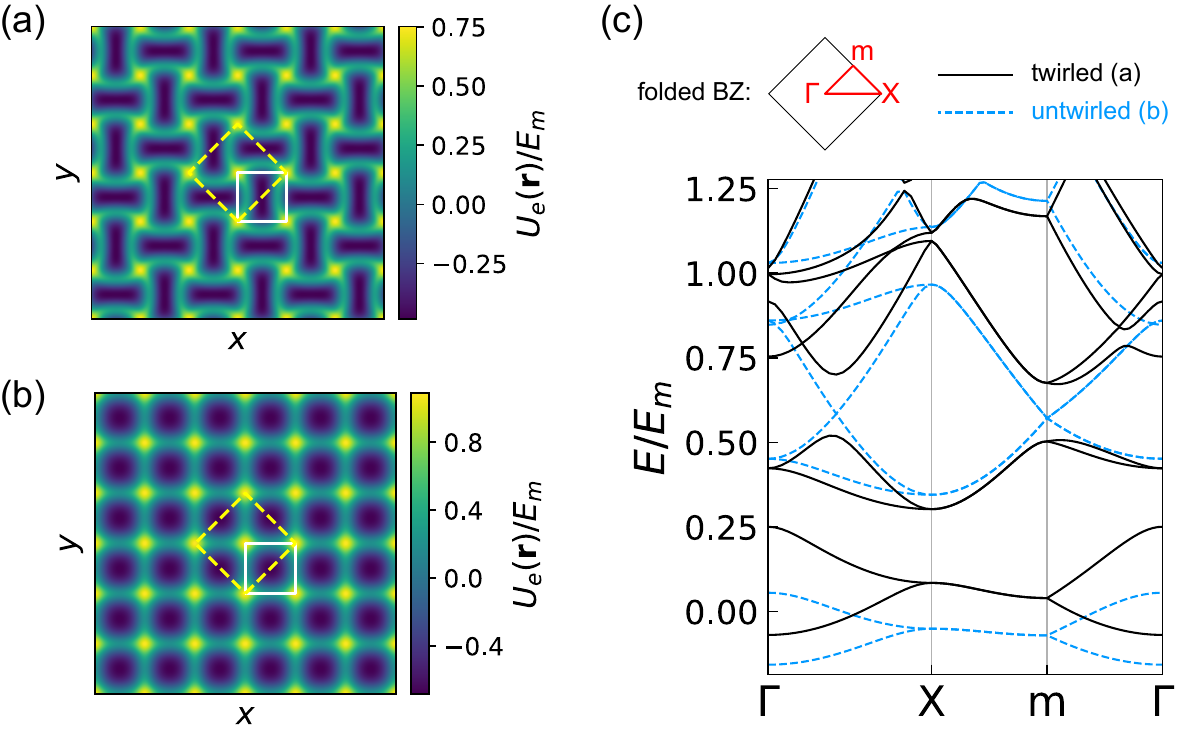}
	\caption{(a), (b) The electron potential $U_e(\rr) = D\nnabla\cdot\uu(\rr)$ obtained under $\epsilon = 0.05$, $\theta = B_{1g} = B_{2g} = 0$, $V_1/\mu = 0.004$ and $D = 10E_m$ via lattice relaxation performed with (a) $\sqrt{2} \times \sqrt{2}$ supercell and (b) the original moir\'e periodicity. The dashed (solid) square is the doubled (original) moir\'e cell.
    (c) The non-interacting moir\'e band structure along the red high-symmetry lines shown on the top. The black solid and blue dashed lines respectively represent the band structure under the electron potential shown in (a) and (b). The bands are spin-degenerate in the non-interacting case.
    The total four-fold degeneracy along $\rm Xm$ is protected by the non-symmorphic and time reversal symmetries of the non-interacting system, and reduces to two-fold spin degeneracy when the magnetism breaks the time reversal symmetry.
    }
	\label{fig:nonint}
\end{figure}

For electron modeling, we assume the general spin-degenerate semiconductor form of electrons or holes in the elastic material, $H_K = \pp^2/2m$, acting on the envelope functions in a single time-reversal-invariant valley with no other flavors than spin.
This description can be appropriate not only for SOC-free systems, but also for centrosymmetric nonmagnetic materials with SOC, where the spin degeneracy is protected by inversion and time-reversal symmetries.
Here $m$ is the effective mass, ranging from $0.01m_e$ to several times $m_e$ in real materials \cite{vurgaftman2001band, chen2020twodimensional} where $m_e$ is the free electron mass.
A typical moir\'e kinetic energy scale is $E_m = \epsilon^2h^2/2ma^2$, which is about 235meV for $m = 0.1m_e$, $a = 4\rm\AA$ and $\epsilon = 0.05$. Relaxed moir\'e pattern generally affects electrons via deformation potential $D\nnabla\cdot\uu(\rr)$, substrate potential $\Delta(\dd(\rr))$ and pseudo-magnetic field $\pp\cdot\AAA(\rr)/2m + H.c.$ \cite{wu2018hubbard, mao2024transfera, zhang2025twistanglea, luskin2026relaxationdriven}.
Pseudo-magnetic field is absent in time-reversal-invariant valleys; in the first-star approximation, the substrate and deformation potentials have similar shapes hence can be effectively merged for our purpose. 
Therefore, we let the non-interacting part of our effective Hamiltonian be $H_0 = H_K + D\nnabla\cdot\uu$, where the effect of $\Delta(\dd(\rr))$ is absorbed into the deformation potential $D$. 
$D$ can be either positive or negative \cite{jin2023highthroughput}, which respectively corresponds to the domain- or junction-centered case in our work. Fig. \ref{fig:nonint}(a) shows an example of deformation potential $D\nnabla\cdot\uu(\rr)$ obtained under $D > 0$, while Fig. \ref{fig:nonint}(b) shows its untwirled counterpart. The $\sqrt{2}\times\sqrt{2}$ periodicity in twirls allows us to obtain the moir\'e band structure in the folded moir\'e Brillouin zone (BZ). One non-interacting example is shown in Fig. \ref{fig:nonint}(c), where we see that the twirls open a significant gap between the 4th and 5th moir\'e bands.

Next, we let the electrons interact via Coulomb potential $C(\rr) = e^2/\tilde\epsilon|\rr|$, where $\tilde\epsilon$ is the dielectric constant of the environment. A typical moir\'e Coulomb energy scale is $C_m = \epsilon e^2/\tilde\epsilon a$, which is about 36meV for $\tilde\epsilon = 5$, $a = 4\rm\AA$ and $\epsilon = 0.05$. The spin-resolved mean-field band structures are obtained via self-consistent Hartree-Fock theory on a discretized BZ grid containing $12\sqrt{2} \times 12\sqrt{2} = 288$ sample points. Detailed formulation is presented in the Supplemental Material \cite{SM2}.
Computer code for our computation is generated by OpenAI Codex with GPT 5.5 model under human-written instruction that contains all mathematical details.

\smallskip
\paragraph{Interacting results.}

\begin{figure}
	\centering
	\includegraphics[width=\columnwidth]{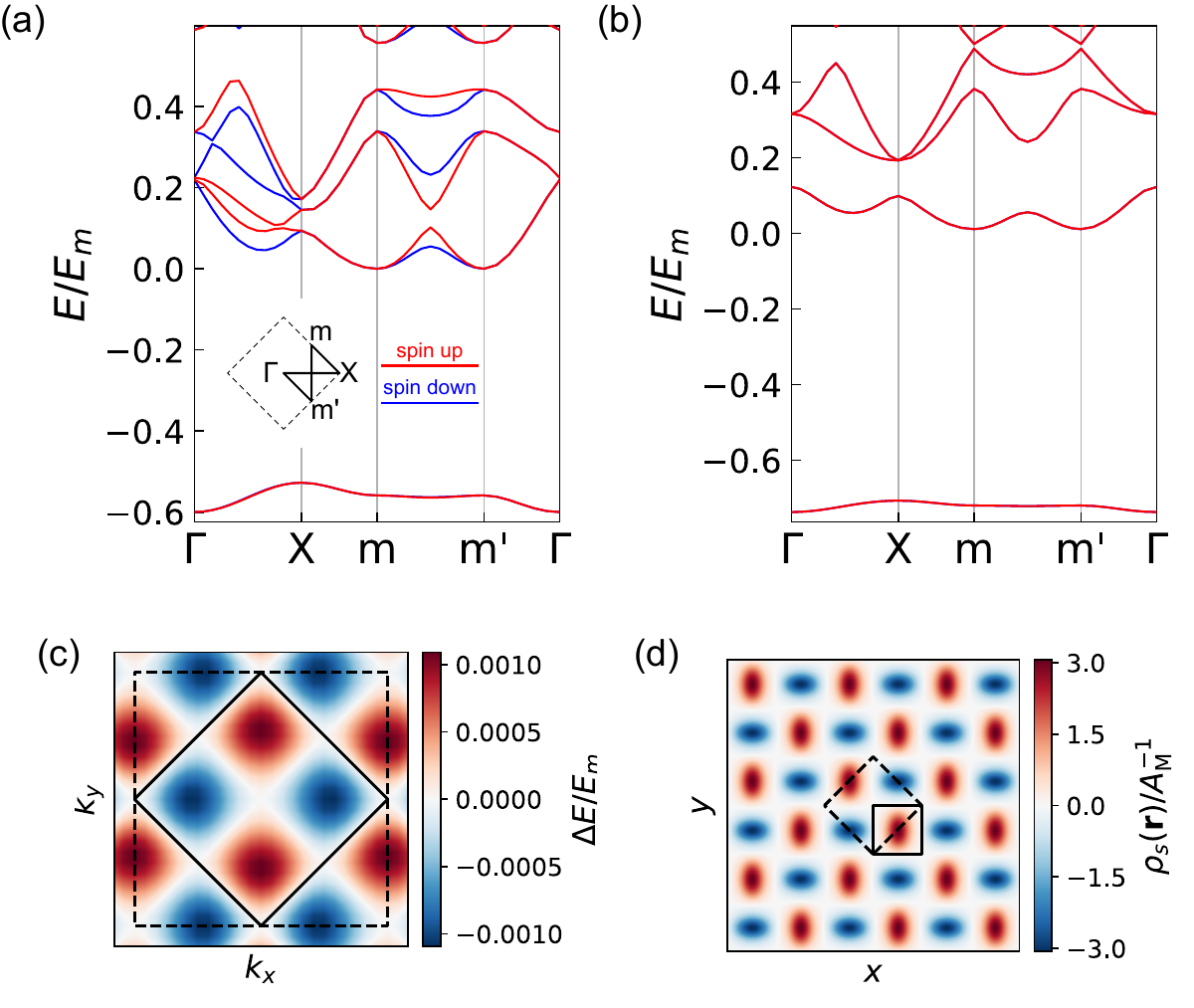}
	\caption{(a) The spin-resolved mean-field moir\'e band structure under $\epsilon = 0.05$, $\theta = B_{1g} = B_{2g} = 0$, $V_1/\mu = 0.004$, $D = 10E_m$, interaction strength $C_m = 0.2E_m$ and filling factor corresponding to one electron per $\sqrt{2} \times \sqrt{2}$ supercell per spin sector (i.e., one filled band in Fig. \ref{fig:nonint}(c) for each spin), plotted along the high-symmetry line shown in the inset. (b) Same as (a), but under the original-moir\'e-periodicity relaxation background illustrated in Fig. \ref{fig:nonint}(b). Here the bands are spin-degenerate. (c) The spin splitting $\Delta E_\kk = E_\kk^\up - E_\kk^\down$ of the lowest band in (a). The solid (dashed) black square is the BZ corresponding to the supercell (the original moir\'e cell). (d) The spin density, $\rho_s(\rr) = \rho_\up(\rr) - \rho_\down(\rr)$, of the lowest band in (a) in units of one per \textit{original} unit cell area $A_\rmM = a^2/\epsilon^2$.}
	\label{fig:domain_centered}
\end{figure}

We present our mean-field results in the $D>0$ case in Fig. \ref{fig:domain_centered}. At band filling corresponding to one electron per $\sqrt{2}\times\sqrt{2}$ supercell in each spin sector, the Coulomb repulsion opens a large correlated insulating gap between the lowest two bands in Fig. \ref{fig:nonint}(c). Fig. \ref{fig:domain_centered}(a) shows spin splitting up to $0.05\sim0.1E_m$ with nodal lines along $\rmX - \rmm$ and $\rmm' - \Gamma$, consistent with $d_{x^2-y^2}$-wave symmetry. The spin splitting is caused by the twirls in the relaxed domain walls, because Fig. \ref{fig:domain_centered}(b) shows that the mean-field moir\'e bands are spin-degenerate under the twirl-free relaxation background. In Fig. \ref{fig:domain_centered}(a), the lowest band has small spin splitting, which is shown in Fig. \ref{fig:domain_centered}(c). The spin density map of the filled band presented in (d) confirms the signature of the schematic lattice model in Fig. \ref{fig:intro}(c). We also show in the Supplemental Material \cite{SM2} that $d$-wave spin splitting occurs not only
at this particular parameter combination, but also in a range of parameters.

The contrast in band splitting between filled and empty bands can be understood with an effective tight-binding picture. In the lowest band, the nearest spin-inequivalent hopping distance between the $s$-like Wannier orbitals is already twice the moir\'e domain size. Because the hopping amplitudes decay exponentially with distance, the effect of spin contrast in this hopping is suppressed.
In higher (or empty) bands, orbitals generally have larger spatial spread (even for the $s$-like orbitals -- because they are lifted up by the Coulomb repulsion with the opposite-spin-occupied states, they strongly hybridize with orbitals with higher angular quantum number), which increases the hopping amplitudes and thus enhances the spin contrast, leading to larger spin splitting.

\begin{figure}
	\centering
	\includegraphics[width=\columnwidth]{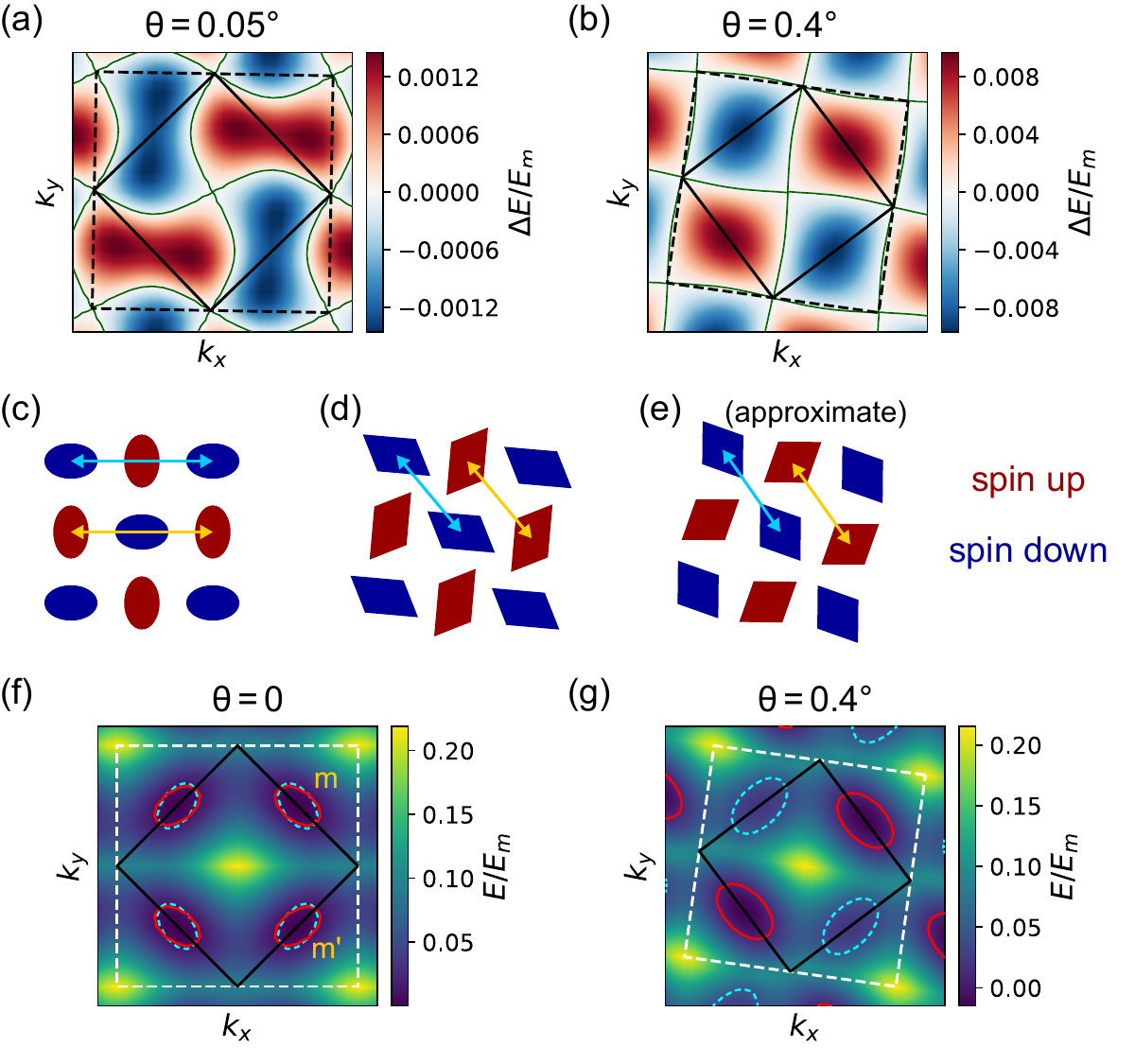}
	\caption{(a), (b) The spin splitting of the lowest mean-field band of a system with the same parameters as the one shown in Fig. \ref{fig:domain_centered}(c), except for a finite twist angle between the elastic material and the substrate illustrated at the top of each panel. Here the curved lines outline the band touching. 
    (c)-(e) Schematics of effective lattice models that describe the lowest bands under (c) $\theta = 0$, (d) relatively small finite $\theta$, and (e) relatively large $\theta$, where the arrows illustrate the nearest spin-inequivalent hopping in the three cases. The shapes are only schematic and reflect the symmetries of the system. The mirror symmetries of (e) are approximate in the physical systems it points to.
    (f), (g) The dispersion of the first spin-up moir\'e band above the correlated insulating gap, in a system with twist angle illustrated on the top of each panel and the rest of the parameters identical to (a), (b). The red solid (cyan dashed) ellipses are the spin-up (spin-down) Fermi surfaces upon doping extra 0.1 electrons per original moir\'e cell, ignoring residual interaction effects of the doping electrons.}
	\label{fig:domain_centered_twisted}
\end{figure}

As a finite twist angle between the elastic material and substrate is turned on, the spin nodal lines in the lowest band become curved due to violation of mirror symmetry, and ultimately rearranges the spin splitting to nearly $d_{xy}$-wave with magnitude enhanced by nearly an order (Figs. \ref{fig:domain_centered_twisted} (a), (b)). Since the twist angle breaks mirror symmetries but preserves $C_{4z}$, the spin degeneracy at $\Gamma$ is maintained; Along any loop around $\Gamma$, the spin splitting has to alternate 4 times between positive and negative, guaranteeing the presence of nodal lines.
Figs. \ref{fig:domain_centered_twisted} (c)-(e) present schematic understanding of the twist angle evolution of band splitting.
Violation of mirror symmetries by the finite twist angle allows for spin-inequivalent hopping between the nearest same-sublattice neighbors, which is forbidden in the zero twist case (Figs. \ref{fig:domain_centered_twisted} (d), (e)).
As the twist angle continues to increase, the spin contrast in the nearest same-sublattice hopping amplitudes grows up to orders of magnitude above the spin contrast in further-neighbor hoppings, thus dominates the spin splitting in the band structure.
Ignoring the spin contrast in further-neighbor hoppings, the system has approximate spin-flipping mirror symmetries $[C_{2\perp}{\parallel}M_{\hat x}]$ and $[C_{2\perp}{\parallel}M_{\hat y}]$ (Fig. \ref{fig:domain_centered_twisted}(e)), leading to approximate $d_{xy}$ splitting. In the junction-centered case ($D<0$), a similar mechanism of spin-splitting enhancement and near-neighbor dominance, driven by strain-induced symmetry lowering and leading to $g$-to-$d$-wave transition \cite{li2025altermagnetism, zhang2026lifshitz, sheoran2026tuning}, is described in End Matter.

One important experimental signature of $d$-wave altermagnetic metals is anisotropic transport properties with spin sectors related via $90^\circ$ rotation \cite{das2024realizing, dou2025anisotropic}.
In our domain-centered system, we consider further doping our correlated insulating states with electrons to enable electron conduction.
To roughly assess the spin-$90^\circ$-related anisotropy effects, we assume that the added electrons do not interact but simply form Fermi surfaces in the mean-field moir\'e bands of the correlated insulator.
Fig. \ref{fig:domain_centered_twisted}(f) shows slightly mismatching Fermi surfaces for spin-up and spin-down.
With a finite twist angle, the energetic symmetry between $\rm m$ and $\rm m'$ mini-valleys in a spin-resolved band is lifted, allowing mini-valley-polarized population in each spin sector (Fig. \ref{fig:domain_centered_twisted}(g)).
The anisotropy of elliptic Fermi surfaces manifest as spin-contrasting anisotropy in effective band mass, suggesting spin-related  $90^\circ$ anisotropic transport. 
Though we do not address them explicitly, we note that correlations can  induce additional exotic behaviors including unconventional superconductivity, charge density wave and strange metal while the correlated insulator is doped \cite{lee2006doping, cao2020strange}.
These phases may cooperate or compete with altermagnetism to create even more interesting phases of matter \cite{bose2024altermagnetism, chen2026pairdensity}.

\smallskip
\paragraph{Discussion.}

We have demonstrated path to realization of correlated insulating altermagnet in moir\'e materials consisting of nonmagnetic layers, within a minimal electronic continuum model setting built upon continuum elasticity model of lattice relaxation. 
The altermagnetic spin-splitting originates from the interplay of antiferromagnetic superexchange in electrons and staggered twirls in square lattice moir\'e domain walls.
According to our results, when the typical moir\'e kinetic energy scale $E_m = \epsilon^2h^2/2ma^2$ is around 0.1 times the deformation potential constant $D$ and the typical moir\'e Coulomb energy scale $C_m = \epsilon e^2/\tilde\epsilon a$ is $\sim 0.2E_m$, the empty mean-field bands have altermagnetic spin splitting $\sim 0.05E_m$.
In real materials $D \sim 1\,\rm eV$ \cite{jin2023highthroughput, chen2020twodimensional}, meaning that when $E_m\sim 100\,\rm meV$ and $C_m\sim20\,\rm meV$ (i.e. when $(\sqrt{m/m_e})(a/\epsilon) \sim 40\,\rm\AA$ and $\tilde\epsilon a/\epsilon \sim 720\,\rm\AA$), the altermagnetic spin splitting is $\sim 5\,\rm meV$, 
which is detectable with high-resolution spin-polarized scanning tunneling microscopy (STM) \cite{balashov2006magnon, brede2023detecting}.
By contrast, since the structural energy saved by twirling is much larger, $\sim 1\rm eV$ \cite{shi2026spontaneous},  the altermagnetic splitting in electron bands is not expected to significantly affect the moir\'e lattice structure.

\begin{table}
\caption{First-principle-based effective band masses and deformation potential constants of monolayer square-lattice M$_2$X from Ref. \cite{chen2020twodimensional}. Note a factor 2 between their and our conventions of the deformation potential constant, and that the sign of $D$ for valence bands is flipped relative to Ref. \cite{chen2020twodimensional}.}
\begin{tabular}{|c|c|c|c|}
    \hline
    material & band & $m/m_e$ & $E_1 = 2D$ (eV)  \\\hline
    Ag$_2$S & conduction & 0.18 & 2.35 \\\hline
    Ag$_2$Se & conduction & 0.19 & 1.97 \\\hline
    \multirow{2}{*}{Au$_2$S} & conduction & 0.08 & 1.95 \\\cline{2-4}
    & valence & 0.12 & 4.33 \\\hline
    \multirow{2}{*}{Au$_2$Se} & conduction & 0.08 & 1.68 \\\cline{2-4}
    & valence & 0.14 & 3.11 \\\hline
\end{tabular}
\label{table:M2X_parameters}
\end{table}

Taking the experimentally realistic dielectric constant $\tilde\epsilon \sim 5$, the parameter regime specified above suggests a moir\'e length scale $a/\epsilon \sim 144\,\rm\AA$ and an effective mass $m \sim 0.08m_e$.
Hence, for actual material realizations we target our search at 2D nonmagnetic centrosymmetric square-lattice semiconductors with light electrons or holes in a single valley.
Plausible candidates include square-lattice M$_2$X where M = Ag, Au and X = S, Se \cite{chen2020twodimensional}, whose parameters are summarized in Table \ref{table:M2X_parameters}.
Additional requirements include: (i) the substrate needs to be an insulator whose bands are out of the way of the altermagnetic moir\'e bands; (ii) the lattice mismatch $\epsilon$ and the structural coupling strength $V_1$ are suitable so that the $\sqrt{2} \times \sqrt{2}$ twirl pattern in the domain wall network is energetically stable.
We leave in-depth investigation of actual material realization for future work.

{\it Acknowledgements:}
We acknowledge computational resources provided by the Texas Advanced Computing Center (TACC).
J.S. and I.M. acknowledge support by the US Department of Energy, Office of Science, Basic Energy Sciences, Materials Sciences and Engineering Division.
M.K. acknowledges the hospitality of the Argonne National Laboratory.
J.S. thanks helpful interaction with R. Fernandez and A. H. MacDonald.
Coding and reference refinement in this work are assisted by OpenAI Codex with GPT 5.5 model.

\bibliography{references}

\section{End Matter}

\begin{figure*}
	\centering
	\includegraphics[width=\textwidth]{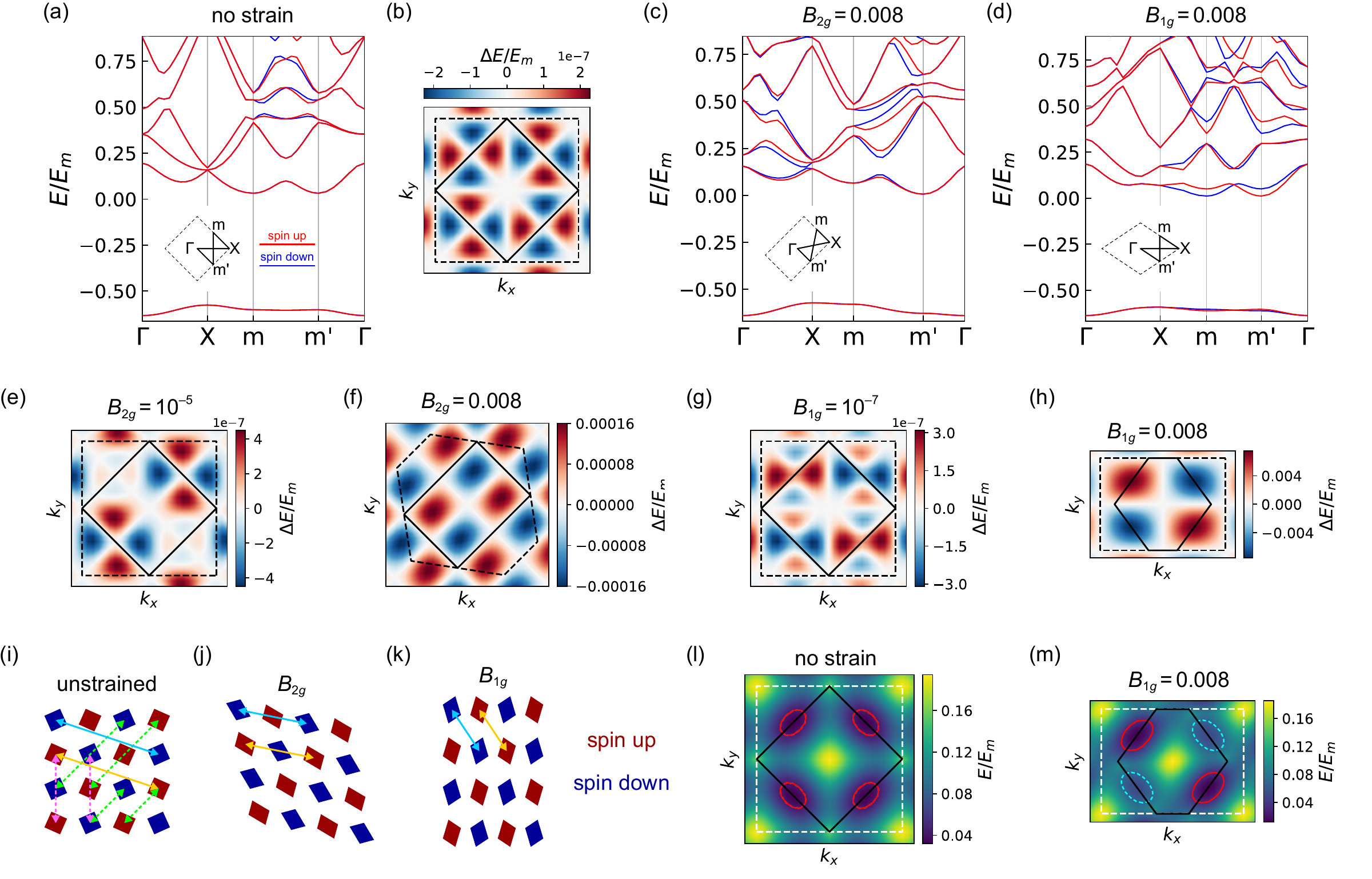}
	\caption{(a) The spin-resolved mean-field moir\'e band structure of the system with the same parameters as in Fig. \ref{fig:domain_centered}(a) except that here $D = -10E_m$ instead of $10E_m$.
    (b) The spin splitting of the lowest band in (a). The solid (dashed) black square is the first BZ of the $\sqrt{2}\times\sqrt{2}$ supercell (the original moir\'e cell). 
    (c), (d) Same as (a) except for nonzero strains marked on the top. The dashed line in the inset of (d) is a direct affine transform from the unstrained moir\'e BZ, which is not the first BZ of the strained moir\'e lattice. However, the vertices of the plotting line are still time-reversal-invariant. 
    (e)-(h) Same as (b) except for nonzero strains marked on the top.
    (i)-(k) Schematics of effective lattice models that describe the low bands with the strain condition indicated on the top. The yellow and blue solid arrows illustrate the nearest spin-inequivalent hopping in each case. The green and pink dashed arrows in (i) are pairs of spin-equivalent hoppings. The shapes are only schematic and reflect the symmetries of the system.
    (l), (m) The dispersion of the first spin-up moir\'e band above the correlated insulating gap, in a system with (l) no strain and (m) $B_{1g} = 0.008$ and the rest of the parameters identical to (a), (b). Again, the red solid (cyan dashed) ellipses are the spin-up (spin-down) Fermi surfaces upon doping extra 0.1 electrons per original moir\'e cell, ignoring residual interaction effects of the doping electrons. In (l), the Fermi surfaces are almost spin-identical because of the negligible $g$-wave splitting.}
	\label{fig:junction_centered}
\end{figure*}

The junction-centered ($D < 0$) mean-field results are presented in Fig. \ref{fig:junction_centered}. In the absence of twist or strain, the bands show $g$-wave splitting, with spin degeneracy along $\rmm'-\Gamma-\rmX-\rmm$, as shown by Figs. \ref{fig:junction_centered} (a), (b). The band splitting is significantly smaller than in the $D > 0$ case presented in Fig. \ref{fig:domain_centered}(a), and is as small as $\sim 10^{-7}E_m$ for the lowest band (Fig. \ref{fig:junction_centered}(b)). Figs. \ref{fig:junction_centered} (c), (d) show that applying strain to the elastic layer can enhance the spin splitting, and that the enhancing effects of $B_{1g}$ strain on the low bands are stronger than those of $B_{2g}$.
Figs. \ref{fig:junction_centered} (e), (f) ((g), (h)) show the evolution of the lowest-band spin splitting with the onset of $B_{2g}$ ($B_{1g}$) strain.
We see that the spin splitting map is already significantly reconstructed under extremely weak ($10^{-5}\sim 10^{-7}$) strains.
As the strain grows, the $g$-wave splitting ultimately transitions to $d_{x^2-y^2}$-wave ($d_{xy}$) in the $B_{2g}$ ($B_{1g}$) case, which is consistent with the symmetry analysis in the \textit{Summary of mechanism} section of the main text.

Like the twist angle evolution in domain-centered systems, here the tininess and strain sensitiveness of $g$-wave splitting can also be understood with schematics of effective lattice presented in Figs. \ref{fig:junction_centered} (i)-(k). In the absence of strain, even the nearest spin-inequivalent hopping spans three moir\'e cells -- all nearer-neighbor hoppings are spin-equivalent by symmetry, as Fig. \ref{fig:junction_centered}(i) illustrates.
Under a $B_{2g}$ ($B_{1g}$) strain, the nearest spin-contrasting hopping becomes the second-neighbor (nearest-neighbor) same-sublattice hopping, as shown in Fig. \ref{fig:junction_centered}(j) (Fig. \ref{fig:junction_centered}(k)), which dominates exponentially over the original spin contrast that gives the $g$-wave band splitting.
Moreover, that the nearest-neighbor spin contrast in hopping is closer under $B_{1g}$ strain than under $B_{2g}$ strain also explains why $B_{1g}$ enhances low-band spin splitting more than $B_{2g}$ does, and why the $g$-wave order is more vulnerable to $B_{1g}$ strain than to $B_{2g}$ strain.

Similarly to the twisted domain-centered system discussed in the main text (Fig. \ref{fig:domain_centered_twisted}(g)), we show in Figs. \ref{fig:junction_centered} (l), (m) that a $B_{1g}$ strain can induce significant spin contrast in Fermi surface and transport anisotropy while the correlated insulating state is further electron-doped. Note that due to violation of 4-fold rotation symmetry by strain, here the spin-resolved transport anisotropies may not be strictly related via $90^\circ$ rotation.

\clearpage

\appendix
\onecolumngrid
\renewcommand\theequation{S\arabic{equation}}
\renewcommand\thefigure{S\arabic{figure}}
\setcounter{equation}{0}
\setcounter{figure}{0}

\section*{Supplemental material for ``Moir\'e-induced altermagnetism from nonmagnetic constituents''}

\subsection{Details of lattice relaxation modeling}

The total lattice energy $E_l[\uu] = \int d^2\rr (U+V)$ is a functional of the local displacement $\uu(\rr)$. Here
\begin{equation}
	U = \frac{\lambda}{2} (\nnabla\cdot\uu)^2 + \mu \lp[ (\partial_x u_x)^2 + (\partial_y u_y)^2 \rp] + \frac{\mu_s}{2} \lp( \partial_x u_y + \partial_y u_x \rp)^2
	\label{eqS:lattice_U}
\end{equation}
is the elastic energy density, where $\lambda$ and $\mu$ are the regular Lam\'e coefficients, and $\mu_s$ is the shear Lam\'e coefficient. A different set of notation, $C_{11}$, $C_{12}$ and $C_{66}$, is often used in past work \cite{chen2020twodimensional, chen2021structural, lv2020metallic} and relates to our notation via
\begin{equation}
	\lambda = C_{12}, \quad \mu = \frac{C_{11} - C_{12}}{2}, \quad \mu_s = C_{66}.
\end{equation}
First-principle values for some materials are listed in Table \ref{tableS:Lame}.

\begin{table}[b]
\caption{Computationally predicted lattice constants and stiffness coefficients of some monolayer square-lattice materials. Here $C_{11}$, $C_{12}$, $C_{66}$ and $\mu$ are in units of $\rm eV/\AA^2$.}
\begin{tabular}{|c|c|c|c|c|c|c|c|c|}
    \hline
    material & reference & lattice constant (\AA) & $C_{11}$ & $C_{12} = \lambda$ & $C_{66} = \mu_s$ & $\mu = (C_{11}-C_{12})/2$ & $\lambda/\mu$ & $\mu_s/\mu$  \\\hline
    s(II)-Cu$_2$S & \cite{chen2020twodimensional} & 5.02 & 2.103 & 0.169 & 1.211 & 0.967 & 0.175 & 1.252 \\\hline
    s(II)-Cu$_2$Se & \cite{chen2020twodimensional} & 4.97 & 2.303 & 0.474 & 1.080 & 0.914 & 0.519 & 1.182 \\\hline
    s(I)-Ag$_2$S & \cite{chen2020twodimensional} & 5.88 & 1.454 & 0.193 & 0.668 & 0.63 & 0.306 & 1.06 \\\hline
    s(I)-Ag$_2$Se &\cite{chen2020twodimensional} & 5.9 & 1.311 & 0.193 & 0.668 & 0.559 & 0.345 & 1.195 \\\hline
    s(I)-Au$_2$S & \cite{chen2020twodimensional} & 5.81 & 1.972 & 0.63 & 0.949 & 0.671 & 0.939 & 1.414 \\\hline
    s(I)-Au$_2$Se & \cite{chen2020twodimensional} & 5.82 & 1.373 & 0.499 & 0.868 & 0.437 & 1.142 & 1.986 \\\hline
    s(I)-Au$_2$Te & \cite{chen2021structural} & 5.85 & 1.464 & 0.372 & 0.768 & 0.546 & 0.681 & 1.407  \\\hline
    s(II)-Au$_2$Te & \cite{chen2021structural} & 5.61 & 2.172 & 1.186 & 0.886 & 0.493 & 2.406 & 1.797 \\\hline
    monolayer FeSe & \cite{lv2020metallic} & 5.32 & 4.955 & 0.328 & 1.218 & 2.313 & 0.142 & 0.527  \\\hline
\end{tabular}
\label{tableS:Lame}
\end{table}

\begin{equation}
	V = V(\dd(\rr)) = \sum_\GG V_\GG e^{i\GG\cdot\dd(\rr)}
	\label{eqS:lattice_V}
\end{equation}
is the coupling energy density to the substrate, which is locally a periodic function of the stacking registry vector $\dd$ with the periodicity of \textit{strained} (if $B_{1g}$ or $B_{2g}$ is nonzero) pre-relaxation lattice of the elastic material. $V_\GG$ is the Fourier coefficient of $V(\dd)$. Given a displacement configuration $\uu(\rr)$, the local stacking configuration is $\dd(\rr) = \mcM\rr + \uu(\rr) = (\mcM_0 + \mcM_B)\rr + \uu(\rr)$, with
\begin{equation}
	\mcM_0 = \begin{pmatrix}
		\epsilon & -\theta  \\  \theta & \epsilon
	\end{pmatrix}, \quad \mcM_B = \begin{pmatrix}
		B_{1g} & B_{2g}  \\  B_{2g} & -B_{1g}
	\end{pmatrix},
\end{equation}
where $\epsilon$ is the lattice mismatch ratio, $\theta$ is the twist angle, and $B_{1g}$ and $B_{2g}$ are respectively the orthirhombic and diagonal area-preserving strains imposed on the elastic layer prior to relaxation. We note that the periodicity of $V(\dd)$ with respect to $\dd$ is affected by $B_{1g}$ and $B_{2g}$ via affine deformation $1 + \mcM_B$, hence specify that $V(\dd) = V_0 \bigl( (1+\mcM_B)^{-1} \dd \bigr)$, and truncate the Fourier components of $V_0(\dd)$ to the first star of reciprocal lattice as in Ref. \cite{shi2026spontaneous}, capturing the substrate coupling strength with a single real parameter $V_1$. In particular,
\begin{equation}
	V_0(\dd) = 2V_1 \lp( \cos\frac{2\pi d_x}{a} + \cos\frac{2\pi d_y}{a} \rp),
\end{equation}
where $a$ is the natural microscopic lattice constant of the elastic layer. The moir\'e cell is spanned by $\AAA_1 = \mcM^{-1}\aaa_1$ and $\AAA_2 = \mcM^{-1}\aaa_2$, where $\aaa_{1,2}$ span the strained microscopic cell, \textit{i.e.}, $\aaa_j = (1+\mcM_B)\aaa_j^0$ with $\aaa_1^0 = (a, 0)$, $\aaa_2^0 = (0, a)$.

\begin{figure}
    \centering
    \includegraphics[width=\textwidth]{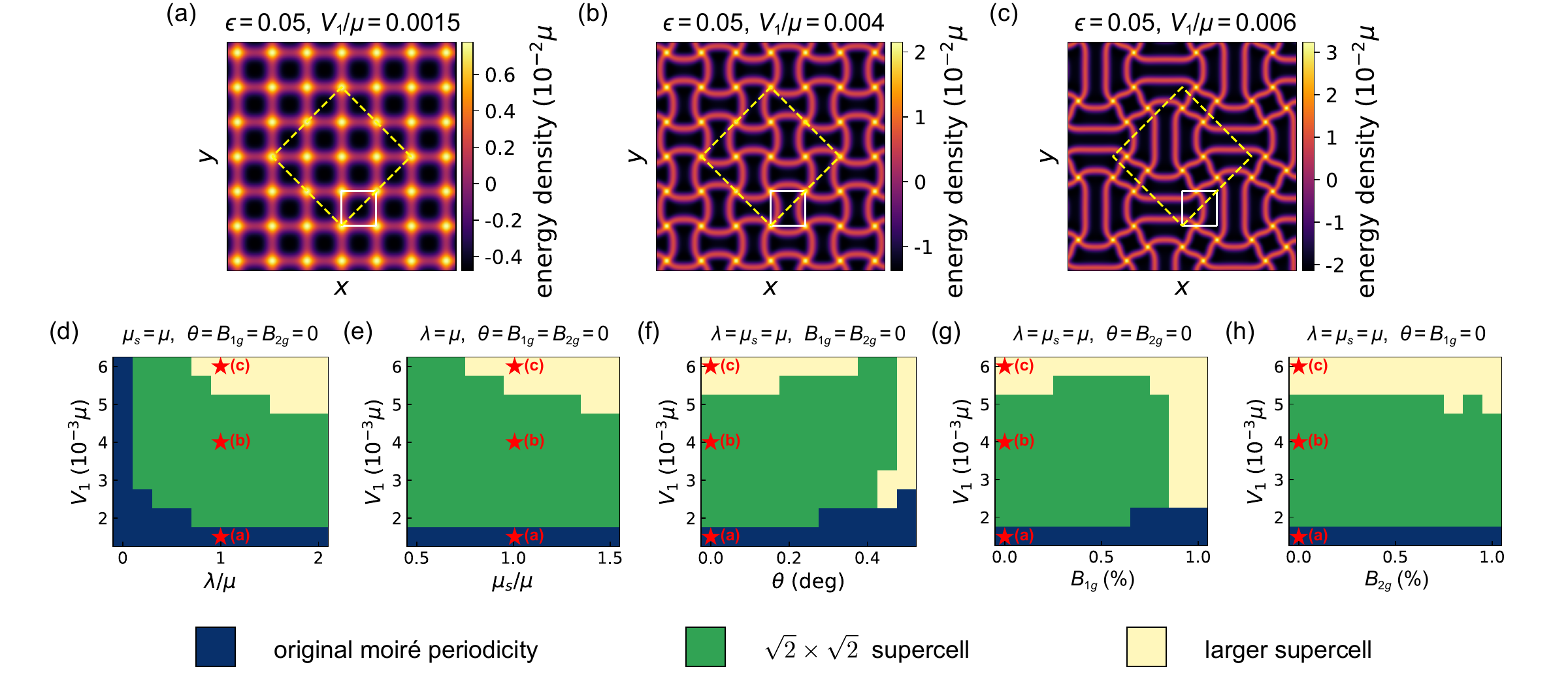}
    \caption{(a)-(c) The local energy density configuration $U(\rr) + V(\dd(\rr))$ of the lowest-energy solution of the relaxation equilibrium equation Eq. (\ref{eqS:relax_equation}), under $\lambda = \mu_s = \mu$, $\epsilon = 0.05$, $\theta = B_{1g} = B_{2g} = 0$ and $V_1/\mu$ indicated on the top of each panel. The solid square is the original moir\'e cell, and the dashed square is the $2\sqrt{2} \times 2\sqrt{2}$ supercell we allow in our relaxation calculation. (d)-(h) Phase diagram of spontaneous moir\'e-translational-symmetry breaking with respect to $V_1$ and various other parameters. In all of them, $\epsilon = 0.05$ and other fixed parameters are labeled on the top of each panel. The red stars mark the positions of the state indicated by (a)-(c) in the phase diagram.}
    \label{figS:relaxation}
\end{figure}

According to Eqs. (\ref{eqS:lattice_U}) and (\ref{eqS:lattice_V}), the variational condition $\delta E_l/\delta\uu = \0$ gives the nonlinear derivative equation
\begin{equation}
	\mu\nnabla^2\uu + (\lambda+\mu) \nnabla(\nnabla\cdot\uu) + (\mu_s-\mu) (\partial_y, \partial_x) (\partial_xu_y + \partial_yu_x) = i\sum_\GG \GG V_\GG e^{i\GG\cdot (\mcM\rr + \uu(\rr))},
	\label{eqS:relax_equation}
\end{equation}
which is numerically solved with the iteration method described in Ref. \cite{shi2026spontaneous} assisted with Anderson acceleration \cite{kelley2022fixed}, allowing for moir\'e translational symmetry breaking up to $2\sqrt{2} \times 2\sqrt{2}$ moir\'e cells. Figs. \ref{figS:relaxation} (a)-(c) show three typical equilibrium states of relaxation that have different periodicities. We see that the system tends to preserve its original moir\'e periodicity under weak $V_1$, while stronger $V_1$ favors moir\'e translational symmetry breaking into larger cells. Figs. \ref{figS:relaxation} (d)-(h) shows that this trend is general under a range of other model parameters including the Lam\'e coefficients, twist angle and strains.

\subsection{Self-consistent Hartree-Fock theory}

We perform Hartree-Fock calculation for electrons under Coulomb interaction and moir\'e potential with $\sqrt{2} \times \sqrt{2}$ supercell, assuming no spin mixing or further translational symmetry breaking by electron interaction. The spin-resolved moir\'e bands are governed by the spin-projected mean-field Bloch Hamiltonians, $H_{\rm MF}^\sigma(\kk) = H_0(\kk) + \Sigma_\rmH + \Sigma_\rmF^\sigma(\kk)$, where $\sigma = \uparrow, \downarrow$ respectively stand for spin up and spin down, $\kk$ is the moir\'e momentum, $\Sigma_\rmH$ and $\Sigma_\rmF^\sigma(\kk)$ are respectively the Hartree and Fock self energy matrices while only the latter can be $\kk$- and spin-dependent. The non-interacting part of the Hamiltonian, $H_0(\kk)$, has matrix elements
\begin{equation}
	H_{0, \gggg\gggg'}(\kk) = \frac{\hbar^2 (\kk+\gggg)^2}{2m} \delta_{\gggg\gggg'} + iD (\gggg-\gggg') \cdot \uu_\gggg,
\end{equation}
where $\gggg$ is a reciprocal lattice vector of the $\sqrt{2} \times \sqrt{2}$ superlattice, and $\uu_\gggg$ is the corresponding Fourier coefficient of $\uu(\rr)$. The self-energy matrices is associated with the spin-resolved one-particle density matrices $\rho^\sigma(\kk)$ by
\begin{equation}
	\Sigma_{\rmH, \gggg\gggg'} = C_{\gggg-\gggg'} (\rho_{\gggg-\gggg'}^\up + \rho_{\gggg-\gggg'}^\down), \quad
	\rho_\gggg^\sigma = \frac{1}{(2\pi)^2} \sum_{\gggg'} \int_{\rm mBZ} d^2\qq \, \rho_{(\gggg+\gggg')\gggg'}^\sigma(\qq)
\end{equation}
and
\begin{equation}
	\Sigma_{\rmF, \gggg\gggg'}^\sigma(\kk) = -\frac{1}{(2\pi)^2} \sum_{\gggg''} \int_{\rm mBZ} d^2\qq\, C_{\pp-\kk+\gggg''} \rho_{(\gggg+\gggg'')(\gggg'+\gggg'')}^\sigma(\qq),
\end{equation}
where $C_\qq = 2\pi e^2/\tilde\epsilon|\qq|$ is the Fourier-transformed Coulomb potential with $C_\0$ set to 0 by working relative to a uniform neutralizing background \cite{reddy2023fractional}. mBZ stands for the folded moir\'e BZ corresponding to the $\sqrt{2} \times \sqrt{2}$ supercell. Given filling factor $\nu^\sigma$ in each spin sector $\sigma$, \textit{i.e.}, the number of doping electrons per $\sqrt{2}\times\sqrt{2}$ \textit{supercell} in each spin sector, the density matrix is in turn determined by the mean-field Hamiltonian via
\begin{equation}
	\rho^\sigma(\kk) = \sum_{n=1}^{\nu^\sigma} \ket{\phi_{n\kk}^\sigma} \bra{\phi_{n\kk}^\sigma},
\end{equation}
where $\ket{\phi_{n\kk}^\sigma}$ is the $n$th normalized eigenvector of $H_{\rm MF}^\sigma(\kk)$. Self-consistency in the relations between $\rho^\sigma$ and $H_{\rm MF}^\sigma$ is reached via iteration assisted with the direct inversion iterative subspace method \cite{pulay1980convergence, pulay1982improved, kudin2002blackbox}.

\begin{figure}
    \centering
    \includegraphics[width=\textwidth]{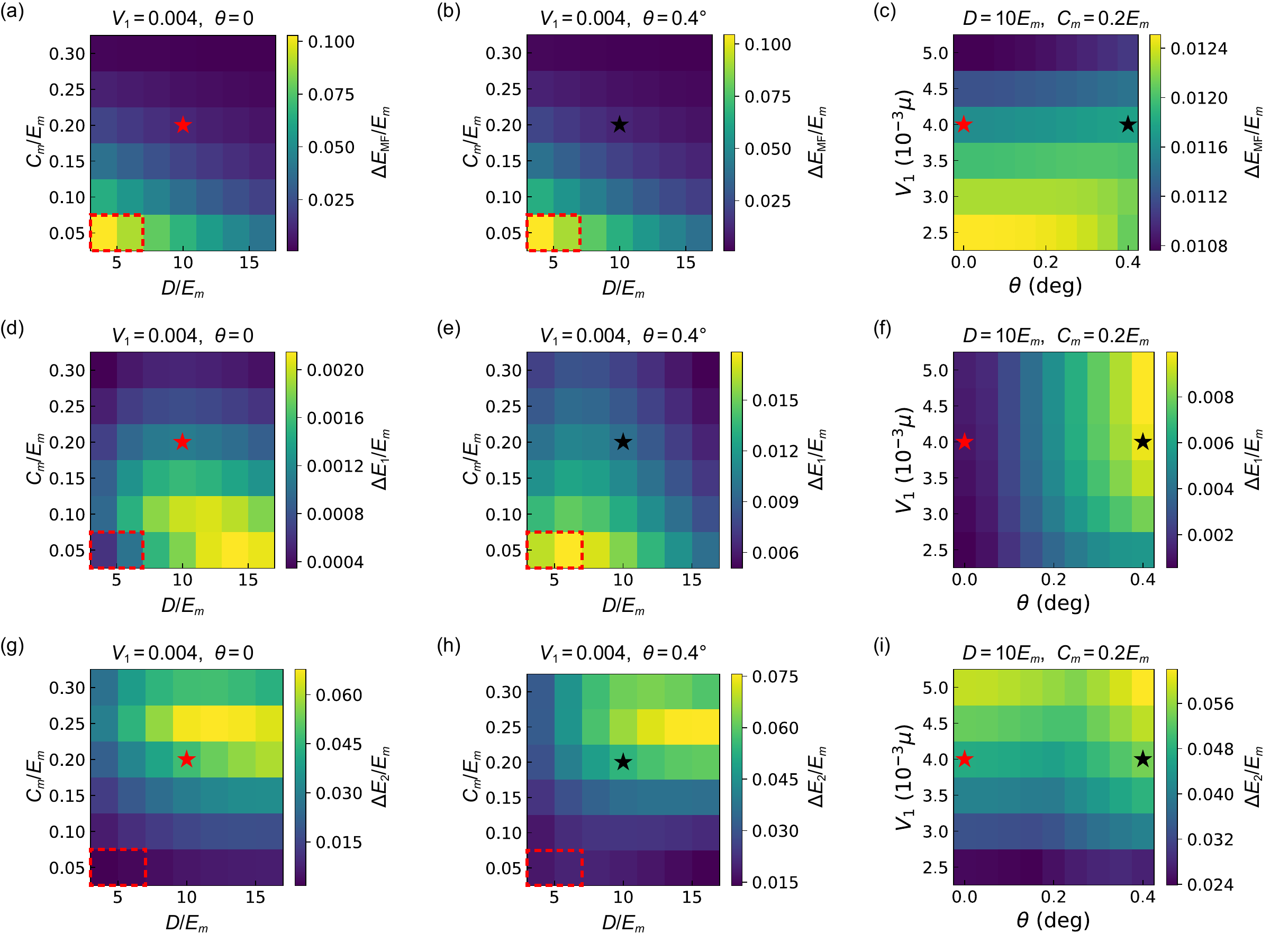}
    \caption{(a)-(c) Parameter dependence of the total mean-field energy difference $\Delta E_{\rm MF} = E_{\rm MF}^{\rm FM} - E_{\rm MF}^{\rm AM}$ between the spin-polarized state (FM, $(\nu^\up, \nu^\down) = (2,0)$) and the spin-compensated state (AM, $(\nu^\up, \nu^\down) = (1,1)$) in the domain-centered system, with fixed and sweeping parameters indicated on the top and axes labels of each panel.
    (d)-(i) Parameter dependence of maximum spin splitting $\Delta E_n = \max_\kk |\Delta E_{n\kk}|$ of (d)-(f) the filled moir\'e band ($n = 1$) and (g)-(i) the first empty moir\'e band ($n = 2$). Here $\epsilon = 0.05$, $\lambda = \mu_s = \mu$ and $B_{1g} = B_{2g} = 0$. The red (black) stars mark the position of the parameter combination under which results are presented in main text Fig. \ref{fig:domain_centered} (Fig. \ref{fig:domain_centered_twisted}(b)). The red dashed boxes mark the regions where the Hartree-Fock correlated gap is indirect and negative, yielding a semimetal state (therefore Hartree-Fock theory may not be accurate in those regimes).}
    \label{figS:HF}
\end{figure}

With the doping density corresponding to one electron per \textit{original} moir\'e cell (i.e., 2 electrons per supercell), we run the Hartree-Fock iterative algorithm under both spin-resolved doping configurations, $(\nu^\up, \nu^\down) = (1, 1)$ and $(2, 0)$, to address possible competition between antiferromagnetic and ferromagnetic couplings \cite{hu2026ferromagnetism} in the electron spins.
For each $(\nu^\up, \nu^\down)$, we start from various independent random initial guesses of the density matrices.
For the convergent solutions, we select the one with the lowest total mean-field energy
\begin{equation}
    E_{\rm MF} = \sum_{\kk\in\rm mBZ} \, \sum_{\sigma = \up, \down} {\rm tr} \lp[ \rho^\sigma(\kk) \lp( H_0(\kk) + \frac{\Sigma_\rmH + \Sigma_\rmF^\sigma(\kk)}{2} \rp) \rp].
\end{equation}
Figs. \ref{figS:HF} (a)-(c) show that under all parameter combinations involved in our Hartree-Fock calculations in this work, the spin-compensated solution ($(\nu^\up, \nu^\down) = (1, 1)$) always has lower $E_{\rm MF}$ than spin-polarized solutions ($(\nu^\up, \nu^\down) = (2, 0)$).

Figs. \ref{figS:HF} (d)-(i) present several maps of band splitting $\Delta E_n = \max_\kk|\Delta E_{n\kk}|$ vs model parameters in our domain-centered system. We see that spin splitting is persistent across wide range in the parameter space. We also confirm that in all the presented parameter regimes the splitting is $d$-wave by checking $\max_\kk|\Delta E_{n\kk} + \Delta E_{nC_4\kk}|$ at each point, which turns out to always be negligible compared to the spin splitting. Here $C_4$ is counterclockwise rotation by $90^\circ$.

\subsection{Scaling invariance}

We consider two moir\'e lattice systems, both with relaxation governed by Eq. (\ref{eqS:relax_equation}) but one of them with primed notations:
\begin{equation}
	\mu'\nnabla^2\uu' + (\lambda'+\mu') \nnabla(\nnabla\cdot\uu') + (\mu_s'-\mu') (\partial_y, \partial_x) (\partial_xu_y' + \partial_yu_x') = i\sum_\GG\GG V_\GG' e^{i\GG\cdot (\mcM'\rr + \uu'(\rr))}.
	\label{eqS:relax_equation_primed}
\end{equation}
Let their parameters relate by $\lambda = \alpha_\mu \lambda$, $\mu' = \alpha_\mu \mu$, $\mu_s' = \alpha_\mu \mu_s$ and $\mcM' = \alpha_\epsilon\mcM$, and let their spatial configurations relate by $\uu'(\alpha_r\rr) = \alpha_a\uu(\rr)$. Eq. (\ref{eqS:relax_equation_primed}) now becomes
\begin{equation}
	\mu\nnabla^2\uu + (\lambda+\mu) \nnabla(\nnabla\cdot\uu) + (\mu_s-\mu) (\partial_y, \partial_x) (\partial_xu_y + \partial_yu_x) = i\frac{\alpha_r^2}{\alpha_\mu\alpha_a} \sum_\GG\GG V_\GG' e^{i\GG\cdot \bigl( \alpha_r\alpha_\epsilon\mcM\rr + \alpha_a\uu(\rr) \bigr)}.
\end{equation}
One can show that if $\alpha_r = \alpha_a/\alpha_\epsilon$ and $V_\GG' = \alpha_\mu\alpha_\epsilon^2 V_{\alpha_a\GG}$, then the equation reproduces Eq. (\ref{eqS:relax_equation}). Therefore, if $\epsilon$ changes, the relative relaxation configuration can be kept invariant by properly scaling other parameters.

For electrons, we use more straight-forward arguments -- the physics is invariant when all moir\'e energy scales vary proportionally. Hence, the amplitude of deformation potential $D\nnabla\cdot\uu(\rr)$ should scale with the moir\'e kinetic energy $E_m = \epsilon^2h^2/2ma^2$; $\nnabla\cdot\uu$ scales with $a/a_M \propto \epsilon$, where $a_M$ is the moir\'e period, hence $D$ scales with $\epsilon/ma^2$. To make the moir\'e-scale Coulomb interaction energy $C_m = |\epsilon|e^2/\tilde\epsilon a$ also scale with $E_m$, the inverse dielectric constant $1/\tilde\epsilon$ should scale with $|\epsilon|/ma$.
We note that the above scaling arguments for both lattice relaxation and electrons also hold even when $\alpha_\epsilon, \alpha_r < 0$ (in which case $D$ changes sign).

We note a subtlety in our scaling argument -- since $\theta$, $B_{1g}$ and $B_{2g}$ need to scale together with $\epsilon$, when either $B_{1g}$ or $B_{2g}$ is nonzero, scaling in $B_{1g}$ and $B_{2g}$ can slightly deviate the shape of the pre-relaxation strained elastic lattice and thus also the shape of the moir\'e periodicity. However, we expect the effects of this deviation on lattice relaxation and moir\'e bands to be small within the continuum regime where $|\epsilon|, |\theta|, |B_{1g}|, |B_{2g}| \ll 1$.

\end{document}